\documentclass[useAMS,usegraphicx]{mn2e}
\usepackage{amssymb}
\usepackage{graphicx}
\usepackage{times}
\usepackage{amssymb}
\usepackage{epstopdf}
\usepackage{graphicx}
\usepackage{footnote}
\usepackage{threeparttable}
\usepackage{float}
\usepackage{pdflscape}

\title[Near Infrared studies of Nova Cephei 2014 and Nova Scorpii 2015]{Near Infrared studies during maximum and early decline of Nova Cephei 2014 and Nova Scorpii 2015}

\author[Srivastava et al.]{Mudit K. Srivastava$^{1}$\thanks{E-mail: mudit@prl.res.in}, N. M. Ashok$^{1}$ \thanks{E-mail: ashok@prl.res.in}, D. P. K. Banerjee$^{1}$\thanks{E-mail: orion@prl.res.in} and D. Sand$^{2}$\thanks{E-mail: david.sand@ttu.edu}
\\
$^{1}$Astronomy and Astrophysics Division, Physical Research Laboratory, Ahmedabad, India \\
$^{2}$Physics Department, Texas Tech University, Lubbock, TX, 79409, USA
}

\begin{document}

\date{Accepted YYYY Month DD.  Received YYYY Month DD; in original form YYYY Month DD}

\maketitle

\label{firstpage}

\begin{abstract}
We present multi-epoch near-infrared photo-spectroscopic observations of Nova Cephei 2014 and Nova  Scorpii 2015,  discovered in outburst on  2014 March 8.79 UT and 2015 February 11.84 UT respectively. Nova Cep 2014 shows the conventional NIR characteristics of a Fe II class nova characterized by strong CI, HI and O I lines, whereas Nova Sco 2015 is shown to belong to the  He/N class with strong He I, HI and OI emission lines. The highlight of the results consists in demonstrating that  Nova Sco 2015 is a symbiotic system containing a giant secondary. Leaving aside the T CrB class of recurrent novae, all of which have giant donors, Nova Sco 2015 is shown to be only the third classical nova to be found with a giant secondary. The evidence for the symbiotic nature is three-fold; first is the presence of a strong decelerative shock accompanying the passage of the nova's ejecta through the giant's wind, second is the  H$\alpha$ excess seen from the system and third is the spectral energy distribution of the secondary  in quiescence typical of a cool late type giant. The evolution of the strength and shape of the emission line profiles shows that the ejecta velocity follows a  power law decay with time ($t^{-1.13 \pm 0.17}$).  A Case B recombination analysis of the H I Brackett lines shows that these lines are affected by optical depth effects for both the novae. Using this analysis we make estimates for both the novae of the  emission measure $n_e^2L$, the electron density $n_e$ and the mass of the ejecta.


\end{abstract}


\section{Introduction}

Nova Cephei 2014  was discovered as a transient by Nishiyama  and Kabashima (2014) at a magnitude of  11.7 on unfiltered CCD frames (limiting magnitude 13.7) taken around 2014 March 8.792 UT. The object was confirmed to be a classical nova by Munari et al. (2014) who obtained a low-resolution spectrogram (range 395-852 nm, 0.21 nm/pixel) on 2014 March 9.792 UT.  The spectrum showed a red continuum with strong emission lines from the Balmer series, O I 777.4 and 844.6 nm, Ca II 849.8 nm,  and Fe II multiplets 42, 48, and 49.  All emission lines showed strong P-Cyg absorptions which were  blue-shifted by 660 km s$^{-1}$ for the Balmer lines, 780 km s$^{-1}$ for the Fe II lines, and 900 km s$^{-1}$ for the O I lines.  The emission lines had a width of about 800 km s$^{-1}$. The intensity of the O I 844.6 nm emission line was seen to be about twice that of O I 777.4 nm, indicating that  there was fluorescent pumping from hydrogen Lyman$\beta$ photons. Photometry on 2014 March 10.094 UT showed a large value of the color B-V = +1.27 which indicated significant reddening consistent with the red slope of the continuum observed in the spectrum. The object's spectrum showed it to be a highly reddened Fe II class nova observed close to maximum brightness. No detailed study of this nova, in any wavelength regime, has been presented till date.
\par
Nova Sco 2015 was discovered as a bright transient on 2015 February 11.8367 UT at an unfiltered CCD magnitude of 8.2 by Tadashi Kojima  using a 150-mm f/2.8 lens $+$ a digital camera (Nakano 2015) . Nothing was visible on a frame from the same camera on Feb. 10.827 UT. (vsnet-alert 18276: \footnote {http://ooruri.kusastro.kyoto-u.ac.jp/mailarchive/vsnet-alert/18276}; AAVSO special notice no. 397 \footnote {http://www.aavso.org/aavso-special-notice-397}). The object was designated PNV J17032620-3504140 on the CBAT Transient Object Confirmation Page (TOCP). An echelle spectrum on 2015 February 13 at 09:38UT (Walter 2015) confirmed that the object was a nova. H$\alpha$ had an equivalent width of -14 nm and full width at half maximum (FHWM) $\sim$2000 km s$^{-1}$. There were symmetrically displaced emission features at about $\pm$ 4500 kms$^{-1}$ which resembled those seen in fast He/N novae. H$\alpha$ and H$\beta$ showed P Cyg absorption features at about -4200,-3200, and -2300 km s$^{-1}$. O I 777 nm and 845 nm were in emission. A strong emission line at 588 nm with a prominent P Cyg absorption was either He I 587 nm or modestly blue shifted Na I. Broad (2000 km s$^{-1}$ FWZI) He I 706 nm emission was possibly also present. Similarly broad emission was seen in the prominent Fe II multiplet 42 lines at 492, 502, and 517 nm, though the first two may have had some He I contribution. The apparently rapid fading and bright possible near-IR counterpart suggested this  was a system with an M giant donor (Walter 2015), like V745 Sco or Nova Sco 2014.

\begin{figure}
\centering
\includegraphics[bb=0 156 295 513,width=3.5in,height=4.5in,clip]{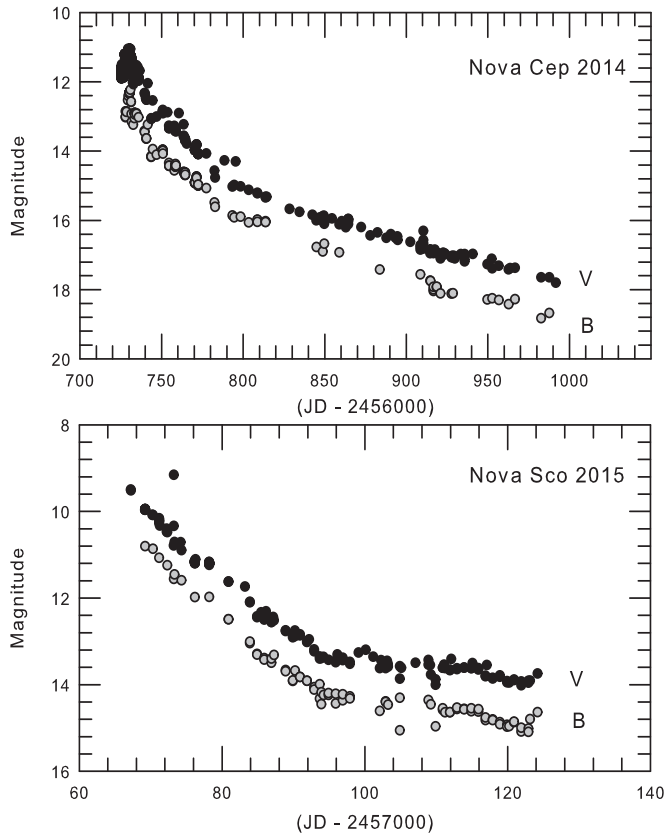}
\caption[]{ The $V$ and $B$ light curves of Nova Cep 2014 and Nova Sco 2015 from  AAVSO data in black and gray symbols respectively.  The outburst dates are  taken as 2014 March 8.792 UT (JD 2456725.2920) and 2015 February 11.8367 UT (JD 2457065.3333) respectively.}
\label{fig_Lightcurves}
\end{figure}


\begin{table}
\centering
\caption{Log of the photometry$^a$ from Mount Abu IR Telescope}
\begin{tabular}{ccccccc}
\hline
Date of      & Days     &       & Magnitudes &         \\
Observation  & since    &   $J$ &  $H$       &    $K$  \\
(UT)         & outburst &       &            &         \\
\hline
\hline
NOVA CEP 2014 & \\
2014 Mar 15.00 &  6.21  &  7.92$\pm$0.02    & 7.39$\pm$0.02  & 7.04$\pm$0.01  \\
2014 Apr 04.98 &  27.19 &  9.31$\pm$0.03    & 8.98$\pm$0.01  & 8.57$\pm$0.02  \\
2014 Apr 06.97 &  29.18 &  9.45$\pm$0.03    & 9.22$\pm$0.02  & 8.78$\pm$0.02  \\
2014 May 05.98 &  58.19 & 10.60$\pm$0.03    & 10.11$\pm$0.02  & 8.90$\pm$0.03 \\
2014 Jun 06.98 &  90.11 & 11.49$\pm$0.06    & 10.74$\pm$0.03  & 9.43$\pm$0.03 \\
2015 Apr 28.00 &  415   & $>$ 15.00       &  $>$ 15.00    &  $>$ 15.00  \\

\hline
\hline

NOVA SCO 2015  & \\
2015 Feb 20.01 &   8.16  &  7.72$\pm$0.24   & 7.14$\pm$0.47  &  6.92$\pm$0.11 \\
2015 Feb 21.01 &   9.16  &  8.22$\pm$0.26   & 7.71$\pm$0.23  &  7.33$\pm$0.21 \\
2015 Mar 05.98 &   22.18 & 10.01$\pm$0.05   & 9.55$\pm$0.16  &  9.09$\pm$0.05 \\
2015 Mar 07.98 &   24.18 & 10.22$\pm$0.09   & 9.76$\pm$0.07  &  9.31$\pm$0.03 \\
2015 Mar 08.95 &   25.18 & 10.28$\pm$0.07   & 9.77$\pm$0.11  &  9.17$\pm$0.26 \\
2015 Mar 19.02 &   35.00 & 10.89$\pm$0.49   &   ----   	     & 	----          \\			
2015 Mar 29.89 &   46.00 & 11.17$\pm$0.20   & 10.50$\pm$0.13 & 10.35$\pm$0.20 \\	
2015 Mar 30.90 &   47.00 & 10.95$\pm$0.08   & 10.38$\pm$0.37 &  9.91$\pm$0.47 \\	

\hline
\end{tabular}
\label{table_ObsPhot}

\begin{list}{}{}
 \item a : Time of outburst is taken as 2014 March 8.792 UT for Nova Cep 2014 and 2015 February 11.8367 UT for Nova Sco 2015.
\end{list}
\end{table}

\begin{table}
\centering
\caption{Log of the spectroscopy observations$^a$}
\begin{tabular}{lcccccc}
\hline
Date of          & Days since                   &&Exposure  time                 \\
Observation      & outburst                      && (s) \\
(UT)             & (days)                        && $(IJ, JH,HK)$                     \\
\hline
\hline
NOVA CEP 2014   &        &&                     \\
 2014 Mar 14.01 &  5.22  &&  ( 300, 240,   240) \\
 2014 Mar 15.02 &  6.23  &&  (----, 180,   180) \\
 2014 Mar 15.98 &  7.19  &&  ( 450, 450,   600) \\
 2014 Mar 20.99 & 12.20  &&  ( 600, 600,  1080) \\
 2014 Apr 05.00 & 27.21  &&  ( 180, 3000,  300) \\
 2014 Apr 07.99 & 30.19  &&  ( 380,  760,  760) \\
\hline
\hline
NOVA SCO 2015 & &\\
 2015 Feb 19.00 &   7.16     &&  (240, 240, 360) \\
 2015 Feb 20.00 &   8.16     &&  (360, 360, 360) \\
 2015 Feb 21.01 &   9.17     &&  (360  360  360) \\
 2015 Feb 25.97 &   14.18    &&  (360, 480, 480) \\
 2015 Feb 26.92 &   15.18    &&  (240, 240, 240) \\
 2015 Mar 06.03 &   22.18    &&  (----, 120, ----)\\
 2015 Mar 09.01 &   25.18    &&  (720, ----, ----)\\
 2015 Mar 23.63$^b$ & 39.80  &&         360      \\
\hline
\hline

\end{tabular}
\label{table_ObsSpec}

\begin{list}{}{}
\item a,b : The spectroscopic observation of Nova Sco 2015 on 2015 March 23.63 UT was made from the IRTF Telescope. The rest of the spectra were obtained  from  Mt. Abu.
\end{list}
\end{table}


\begin{figure*}
\centering
\includegraphics[angle=0,width=1.10\textwidth]{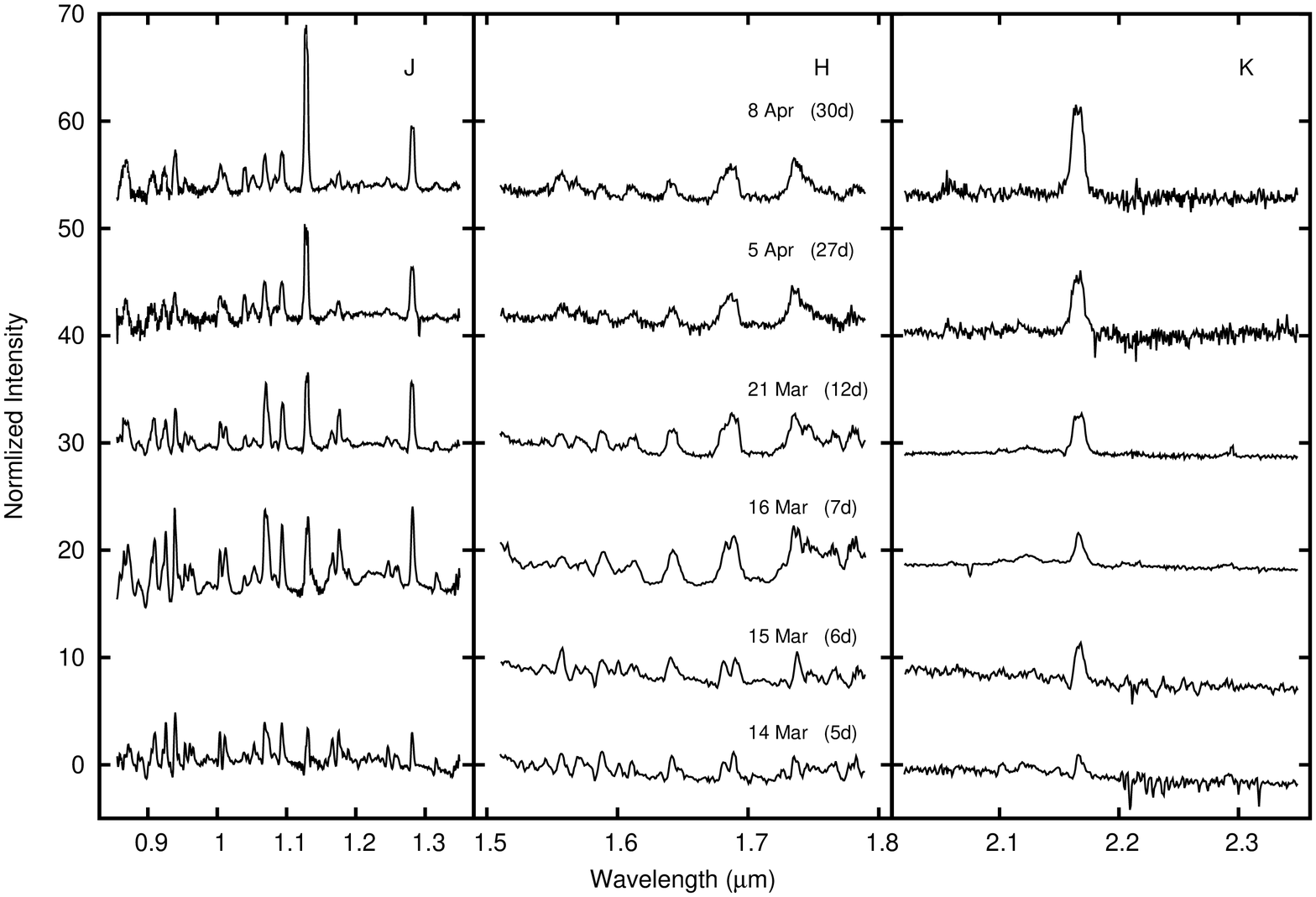}
\caption[]{
J, H and K bands spectra of Nova Cep 2014 normalized to unity at the band centres of 1.25 $\mu$m, 1.65 $\mu$m and 2.20 $\mu$m respectively and then offset for clarity. Days after the outburst are also indicated in parentheses.
}
\label{NCep14_spec_JHK}
\end{figure*}

\par
Early X-ray and radio observations of Nova Sco 2015 by Nelson et al (2015) implicated strong shocks against a red giant wind. Their   observations of Nova Sco 2015 were carried out at X-ray, UV and radio wavelengths. The X-ray observations were carried out with the Swift satellite between 2015 February 15.5 and 16.3 UT (about 4 days after discovery). An X-ray source was clearly detected at the position of the nova.  The spectrum was hard and could be modeled as a highly absorbed, hot thermal plasma (N(H) $\sim$ 6$\pm$1 x 10$^{22}$ cm$^{-2}$; kT greater than 41 keV). However, a significant excess of counts over the model prediction was observed between 1 and 2 keV, possibly indicating the presence of a second, softer emission component. Nelson et al. (2015) also observed Nova Sco 2015 at radio wavelengths with the Karl G. Jansky Very Large Array (VLA) on 2015 February 14.5, approximately 3 days after optical discovery. The nova was detected at frequencies from 4.55 to 36.5 GHz with a spectrum typical of non-thermal synchrotron emission (spectral index between -0.7 and -0.9). The presence of hard, absorbed X-rays and synchrotron radio emission at this early stage of the outburst suggested that the nova-producing white dwarf was embedded within the wind of a red-giant companion, with collisions between the ejecta and this wind shock-heating plasma and accelerating particles (as in, e.g. RS Oph, V407 Cyg and V745 Sco (Banerjee et al. 2009, Munari et al. 2011, Banerjee et al. 2014). This interpretation is supported by our NIR observations.
\par
In this paper we present our NIR spectroscopic and photometric observations of Nova Cep 2014 and of Nova Sco 2015, preliminary reports of which were made in Ashok et al. (2014) and Srivastava et al.(2015). The observations of Nova Cep 2014 span 9 epochs covering 5 to 90 days after the outburst and the observations of Nova Sco 2015 span 11 epochs covering 7 to 47 days after the outburst. We present the observations in Section~\ref{sec_observations}. The analysis and results for Nova Cep 2014 and Nova Sco 2015 are described in Section~\ref{sec_NCep_Results} and Section~\ref{sec_NSco_Results} respectively.

\section{Observations}
\label{sec_observations}

Near-IR spectroscopy in the 0.85 to 2.4 $\mu$m region at resolution $R$ $\sim$ 1000 was carried out with the 1.2m telescope of the Mount Abu Infrared Observatory using the Near-Infrared Camera/Spectrograph (NICS) equipped with a 1024x1024 HgCdTe Hawaii array. Spectra were recorded with the star dithered to two positions along the slit with one or more spectra being recorded in both of these positions. The co-added spectra in the respective dithered positions were subtracted from each other to remove sky and dark contributions. The spectra from these sky-subtracted images were extracted using IRAF tasks and wavelength calibrated using a combination of OH sky lines and telluric lines that register with the stellar spectra. To remove telluric lines from the target's spectra, it was ratioed with the spectra of a standard star (SAO 18998 spectral type A2IV in case of Nova Cep 2015 and SAO 206599, spectral type A0/A1V in the case of Nova Sco 2015) from whose spectra the hydrogen Paschen and Brackett absorption lines had been removed. The spectra were finally multiplied by a blackbody at the effective temperature of the standard star to yield the resultant spectra. All spectra were covered in three settings of the grating that cover the $IJ, JH$ and $HK$ regions separately.
\par
A spectrum of Nova Sco 2015 was obtained using the 3m IRTF telescope on 2015 March 23.625UT covering the 0.8 to 2.5 $\mu$m region. This spectrum was obtained using SpeX (Rayner et al. 2003) in the cross-dispersed mode using the $0.3" \times 15"$ slit ($R = 2000$) and a total integration time of 360s. The SpeX data were reduced and calibrated using the Spextool software (Cushing et al. 2004), and corrections for telluric absorption were performed using the IDL tool xtellcor (Vacca et al. 2003). The log of the observations are given in Tables~\ref{table_ObsPhot} and ~\ref{table_ObsSpec}.


\begin{figure*}
\centering
\includegraphics[bb=3 5 397 292,width=7in,height=4in,clip]{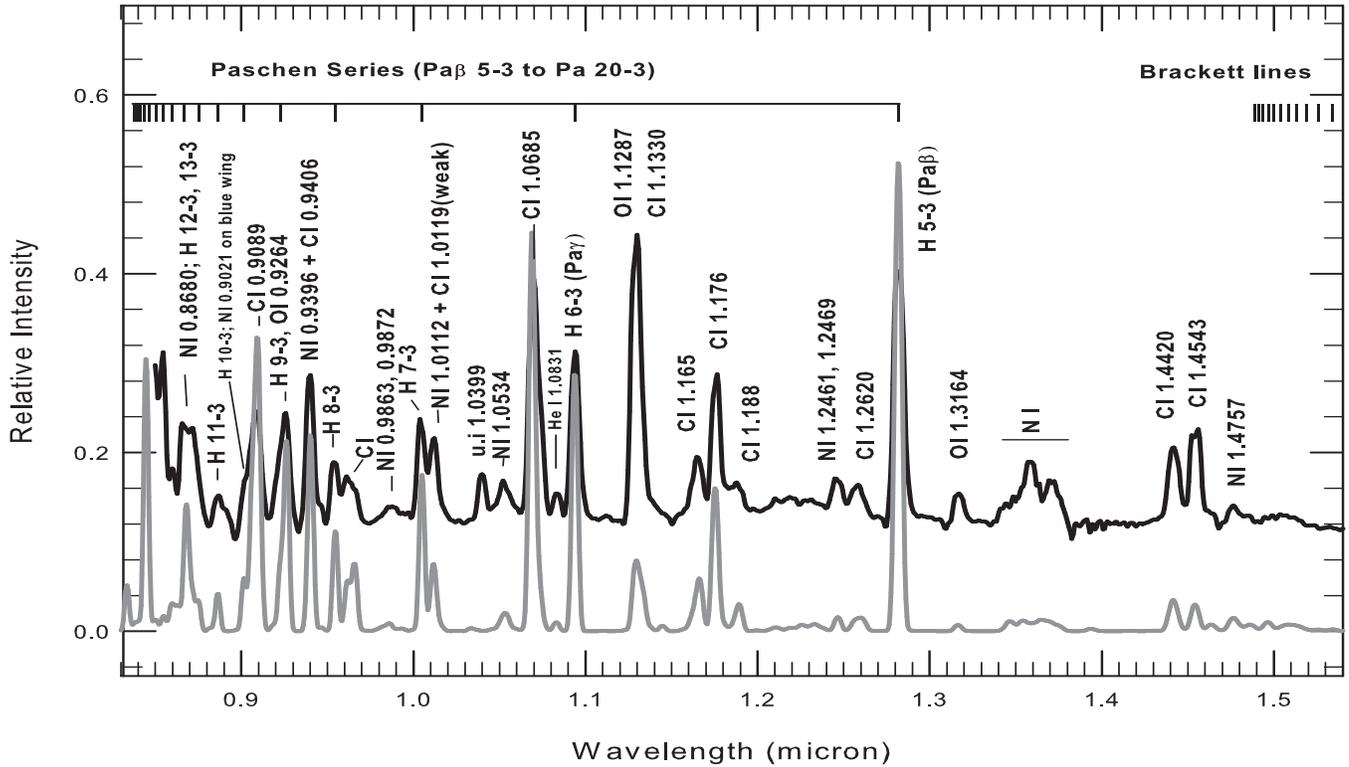}
\caption[]{Line identification of Nova Cep 2014 in the 0.85 - 1.5 $\mu$m region. The observed  spectrum of Nova Cep 2014 of 20th March is shown in black and a typical model LTE spectrum is shown in  gray below to help in line identification. Further details are given in the Appendix.}
\label{fig_NCep_LineID}
\end{figure*}



\section{Results on Nova Cep 2014}
\label{sec_NCep_Results}

\subsection{Light curve, extinction and distance of Nova Cep 2014}
\label{subsec_Ncep_Lightcurve}

Fig~\ref{fig_Lightcurves} shows the $V$ and $B$ band light curves of the Nova Cep 2014 using data from American Association of Variable Star Observers (AAVSO). The nova showed a climb to maximum that lasted for 5 days before peaking at $\sim$ 11.05 mag in $V$ on 2014 March 13.9198 (JD2456730.4198). From the light curve we determine t$_{2}$ and t$_{3}$ - the time for the brightness to decline by 2 and 3 magnitudes respectively from maxima - to be 22$\pm$ 2d and 42 $\pm$ 1d thereby putting it in the fast speed class. The observed $(B-V)$ values at maximum  and at t$_{2}$  equal 1.18 and 0.9 respectively in contrast to the expected values of 0.23 $\pm$ 0.06 and -0.02 $\pm$ 0.04 respectively at these epochs (van den Bergh \& Younger 1987). The large values of $(B-V)$ imply considerable reddening; the excess $E(B-V)$ values are equal to 0.95 and 0.92 respectively. We adopt a mean value of 0.935 for the reddening and thus an extinction $A_{V} = 3.09 \times E(B-V)= 2.9$. For $t_{2}=22 \pm 2d$, the MMRD relation of della Valle \& Livio (1995) gives M$_{V}$ = -7.84 $\pm$ 0.5 which implies a distance to the nova of 15.8 $\pm$ 4 kpc. Similarly, the MMRD relations for t$_{2}$ and t$_{3}$ by Downes \& Duerbeck (2000) yield similar M$_{V}$ magnitudes of -7.90 $\pm$ 0.66 and -7.87 $\pm$ 0.92 respectively. These translate to a mean  distance estimate of 17.2 $\pm$ 7 kpc. Clearly, a large distance to the nova is suggested. The possibility is unlikely that the distance estimate is being boosted up because of a low choice of extinction $A$ in the distance-modulus relation $m - M = 5logd -5 + A$ . The extinction of 2.9 that we have used is close to the total galactic extinction of 2.995 mag  in the direction of the nova as estimated by Schlafly \& Finkbeiner (2011) from dust extinction maps. Our choice of A$_{V}$ is also consistent with the extinction modeling of Marshall et al (2006) who find that the extinction A$_{V}$ rapidly rises, in the direction of the nova, to 3.36 $\pm$ 0.33 by a distance of 1.23 kpc and remains at this value for larger distances.


\subsection{General properties of the spectra and a Case B recombination analysis}
\label{subsec_NCep2014_SpecProp}

The J, H and K band spectra for Nova Cep 2014 are shown in Fig~\ref{NCep14_spec_JHK}. These show that the outburst and evolution of Nova Cep 2014 was that of a conventional Fe II class nova. The spectra of such novae, in the near-IR, are characterized by strong HI lines of the Paschen and Brackett series but what differentiates them from the He/N class is the presence of several strong CI lines seen around maximum and during the early decline (Banerjee \& Ashok, 2012). These CI lines are all prominently seen in the spectrum of Nova Cep 2014 (examples being CI 1.0685, 1.165, 1.176 $\mu$m and a strong blend of CI lines in the region 1.73 to 1.78 $\mu$m). The detailed identification of the lines is presented in Fig~\ref{fig_NCep_LineID} and is discussed in the Appendix and Table~\ref{table_LineList}. P-Cygni absorption components are seen in many of the lines in the spectra taken during 2014 March.  The line profile widths do not vary much over time; for e.g. the FWHM of the Paschen $\beta$ 1.2818 $\mu$m line changes from $\sim$1200 km$s^{-1}$ to $\sim$1500 km$s^{-1}$ between the epochs 2014 March 14 (5.22d) to 2014 April 7 (29.20d).

\begin{figure}
\centering
\includegraphics[angle=0,width=0.50\textwidth]{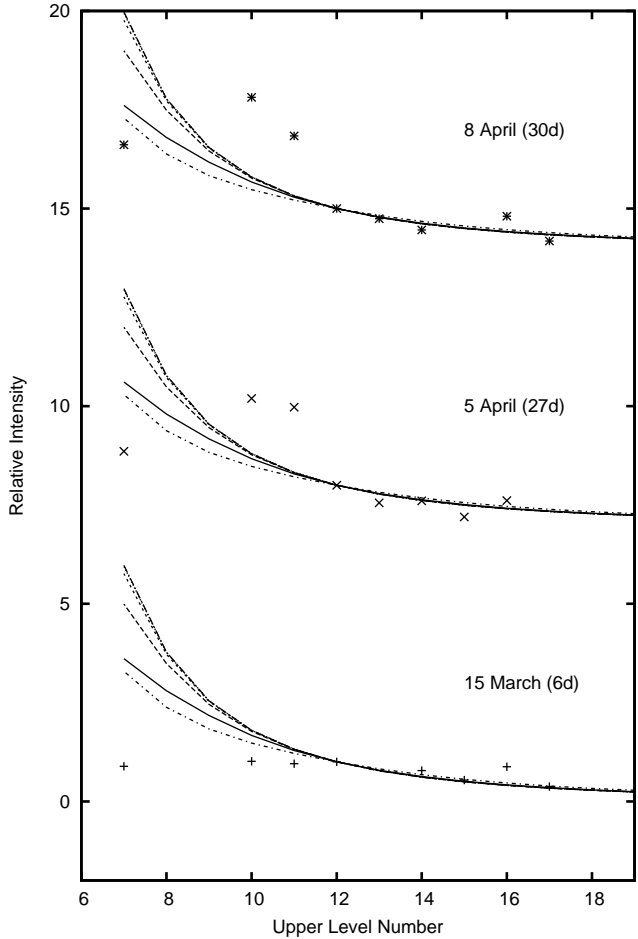}
\caption[]{Recombination Case B analysis for the Hydrogen Brackett series of lines for Nova Cep 2014. The X axis is upper level principle quantum number of the line transition. The line intensities (data points) are normalized with respect to Br 12 line strength and offsets are added for the sake of clarity. The model predictions for Case B analysis are also shown for temperature = 10000 K and electron number densities of
10${^{\rm 9}}$ cm${^{\rm -3}}$ (Small dash-dotted lines),
10${^{\rm 10}}$ cm${^{\rm -3}}$ (solid lines),
10${^{\rm 11}}$ cm${^{\rm -3}}$ (big dash lines),
10${^{\rm 12}}$ cm${^{\rm -3}}$ (small dash lines),
10${^{\rm 13}}$ cm${^{\rm -3}}$ (dotted lines), and
10${^{\rm 14}}$ cm${^{\rm -3}}$ (dash-dotted lines).
}
\label{NCep14_RecombAnalysis}
\end{figure}

\par
We do not find any evidence of dust formation in this nova, as manifested by an IR excess, during the early decline stage. To check whether  dust formation may have occurred later, photometry was done recently (2015 April 28, JD 2457140.5). However the nova was not detected in any of the J, H or K bands. The limiting magnitudes of our observations in J, H and K band are $\sim$15.0. This, taken in conjunction with the latest V magnitudes of 18.943 on 2015 April 6.87 UT (JD 2457119.36553) from the AAVSO database, yields  $(V-K)$ $<$ 3.9. The small value of the $(V-K)$ color  indicates that dust formation is very unlikely to have occurred.
\par
A recombination case B analysis was done, but only  for selected dates of 2014 March 15.02, April 05.00 and  April 07.99 (i.e. 6, 27 and 30 days after the outburst) when contemporaneous photometric observations were available for flux calibrating the spectra.  The measured line fluxes for the H I Brackett lines are given in Table~\ref{table_LineLuminosity_NCep2014} and Fig~\ref{NCep14_RecombAnalysis} shows the Brackett line strengths  with respect to Br 12 set to unity. We find that the line fluxes do not match predicted Case B recombination values. In particular, it is seen from Fig~\ref{NCep14_RecombAnalysis} that the Br7 (Br$\gamma$) line strength is significantly lower than  the predicted values of Storey \& Hummer (1995). Though expected to be stronger than the other Br series of lines, it is found to be significantly weaker than for e.g Br10 and Br11. Such behavior is expected in the early phase of outbursts signifying that Br$\gamma$ is optically thick and so possibly are the other Br lines. Such  optical depth effects in the Brackett lines are also seen in several other novae systems e.g. Nova Oph 1998 (Lynch et al. 2000), V2491 Cyg and V597 Pup (Naik at al. 2009), RS Oph (Banerjee et al. 2009) and T Pyx (Joshi et al. 2014).


\begin{table}
\centering
\caption{List of emission line flux values at different epochs for Nova Cep 2014}
\begin{tabular}{lccc}

\hline\\

Emission Line & & Integrated Line Flux    &    \\
and Wavelength& & at Days after Outburst  &    \\
($\mu$m)      & & ($10^{-20}$ Watt/cm$^2$) &    \\
\hline
            & 6.23d & 27.21d  & 30.19d \\
\hline
\hline

Br17 1.5439 &	35.0 &	---- &  7.44  \\
Br16 1.5556 &	81.4 &	28.1 &	33.5  \\
Br15 1.5701 &   50.1 &	9.04 &	----- \\
Br14 1.5881 &	72.5 &	27.9 &	19.0  \\
Br13 1.6109 &   ---- &	25.5 &	30.8  \\
Br12 1.6407 &	92.8 &	46.0 &	41.5  \\
Br11 1.6807 &	88.5 &	137.0&	118.0 \\
Br10 1.7362 &	94.1 &	147.0&	158.0 \\
Br7  2.1655 &	82.4 &	85.5 &	108.0 \\

\hline
\hline
\end{tabular}
\label{table_LineLuminosity_NCep2014}
\end{table}


\begin{figure*}
\centering
\includegraphics[angle=0,width=1.10\textwidth]{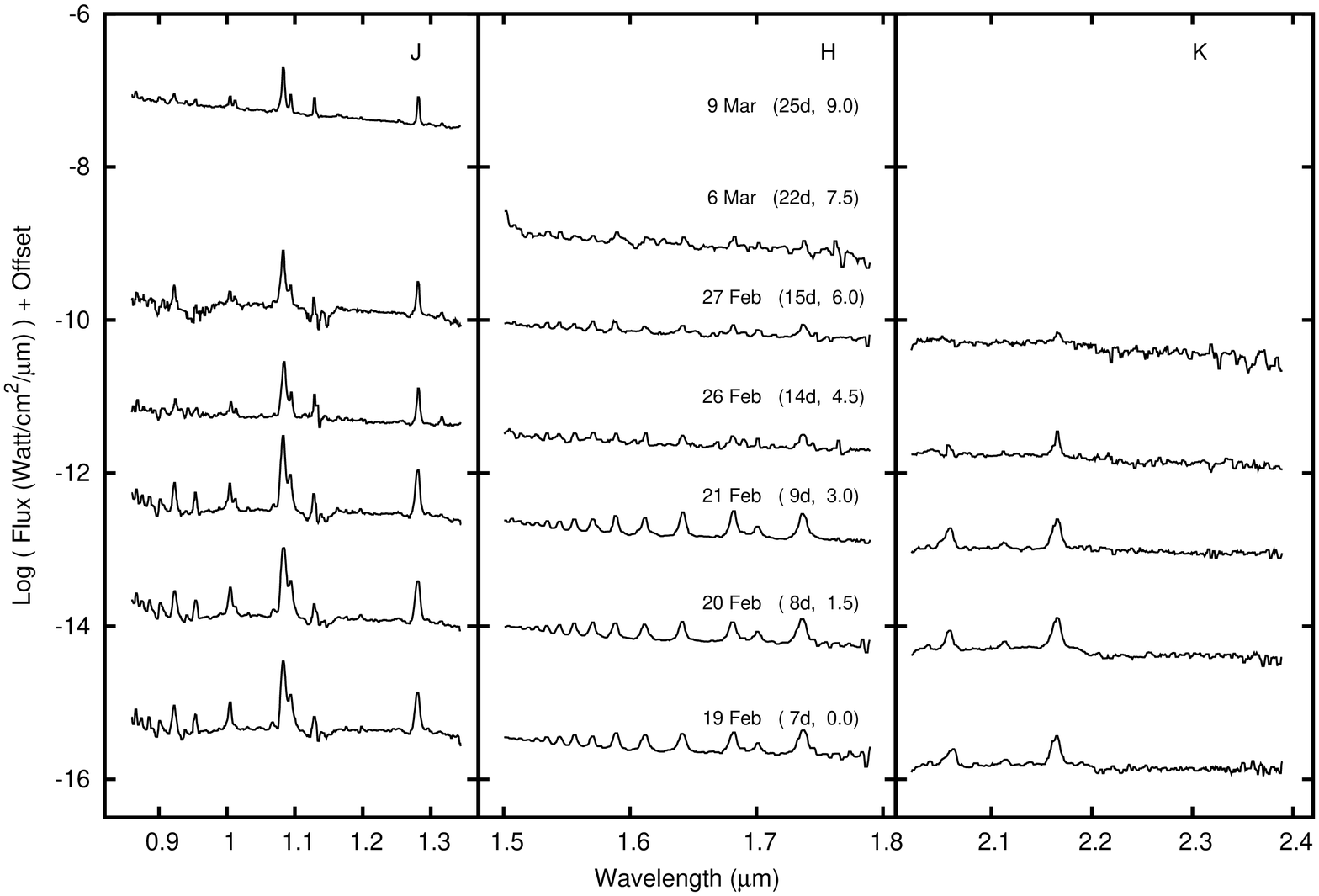}
\caption[]{Flux Calibrated J, H and K bands spectra of Nova Sco 2015 with the days after outburst and offset indicated in the parentheses. Offsets are applied to the intensities for sake of clarity.}

\label{NSco15_spec_JHK}
\end{figure*}

\par
Although the lines are optically thick,  we can estimate the emission measure $n_e^2L$ of the ejecta following the opacity data given by Hummer \& Storey (1987) and Storey \& Hummer (1995) and using the fact that Br$\gamma$ line is found to be optically thick. The optical depth at line-center $\tau_{n,n'}$ is given by $\tau = n_e n_i \Omega(n,n') L$, where $n_e$, $n_i$, $L$ and $\Omega(n,n')$ are the electron number density, ion number density, path length and opacity corresponding to the transition from upper level $n$ to lower level $n'$, respectively. Further,  the opacity factor $\Omega(n,n')$ does not vary significantly within the density or temperature range that is expected to prevail in the ejecta. For e.g., from Storey \& Hummer 1995, the opacity $\Omega(n,n')$ for Br$\gamma$ line for the temperature $T_e = 10000$ K, and number densities $10^9$ to $10^{13}$ cm$^{-3}$ vary only between 1.3 $\times$ 10$^{-34}$ to 7.46 $\times$ 10$^{-34}$. We will assume that the densities in the early stage of the nova outburst are  high and lie in the in the above range of $10^9$ to $10^{13}$ cm$^{-3}$. As $\tau =n_e n_i \Omega(n,n') L >> 1 $ the emission measure $n_e^2L$ for above values is estimated to  be in the range of $1.3 \times 10^{33}$ to $7.7 \times 10^{33}$ cm$^{-5}$.
\par
We constrain the electron density by taking $L$ as the kinematical distance $v \times t$ traveled by the ejecta where $v$ is the velocity of ejecta and $t$ is the  time after outburst. We consider a typical  ejecta velocity of  $v$ $\sim$ 1000 kms$^{-1}$ as measured from half the FWZI of the Pa$\beta$ 1.2818 $\mu$m line and  $t$ to range from 6 to 30 days. With  the constraints that $\tau (Br\gamma)= n_en_i \Omega(n,n') L > 1$, the lower limit on electron density $n_e$ is found to be  in the range $2.2\times 10^9$ cm$^{-3}$ to  $1.1 \times 10^{10}$ cm$^{-3}$ (assuming $n_e = n_i$).  It should be noted that these derived lower limits are likely to be smaller than the actual $n_e$ values because $\tau$ (Br$\gamma$) can be considerably $> 1$.
\par
The density in the nova ejecta remains significantly high over the entire duration of our observations. Lynch et al. (2000) showed that high densities of $10^{10}$ cm$^{-3}$ or more tend to thermalize the level populations through collisions and thereby bring about deviations from Case B predictions. The same has been observed here in HI lines.
\par
The gas mass of the ejecta may be estimated by $M$ = $\epsilon$$V$$n_e$$m_H$ where $V$ is the volume ( = 4/3$\pi$$L^{3}$), $\epsilon$ is the volume filling factor  and $m_H$ is the proton mass. For $L$ varying between the distance traversed in 6d to 30d and the corresponding lower limits on $n_e$ as estimated above, the mass $M$ varies between ($5.4 \times 10^{-6}$ -- $1.3 \times 10^{-4}$)$\epsilon$  M$_{\odot}$. This is a wide range and the mass is  poorly constrained but nevertheless the mass range is consistent with the typical ejecta masses estimated in novae of $10^{-4}$ to $10^{-6}$ M$_{\odot}$.


\section{Results on Nova Sco 2015}
\label{sec_NSco_Results}

\subsection{Light curve, extinction and distance of Nova Sco 2015}
\label{NSco2015_distance}
\par
The V and B band light curves are shown in the lower panel of Fig~\ref{fig_Lightcurves} using the data from AAVSO. The nova showed a monotonic decline and we determine t$_{2}$ and t$_{3}$ values of 14$\pm$2 d and 19$\pm$1 d which puts the Nova Sco 2015 in the fast speed class similar to Nova Cep 2014 discussed earlier in subsection~\ref{subsec_Ncep_Lightcurve}. The observed (B-V) value near the optical maximum and t$_{2}$ are 0.87 and 0.79 respectively. Comparing these values with the expected values of 0.23 $\pm$ 0.06 and -0.02 $\pm$ 0.04 respectively at these epochs from van den Bergh \& Younger (1987), we get an average value of 0.72 for the color excess E(B-V) and interstellar extinction $A_V=2.23$. By using the MMRD relation of della Valle \& Livio (1995) we get $M_V= -8.44 \pm 0.14$ for $t_{2}=14$ d which implies a distance of $13.7 \pm 0.4$ kpc for $A_V=2.23$. Similarly by using the MMRD relations for t$_{2}$ and t$_{3}$ of Downes \& Duerbeck (2000) we get  M$_{V}$ values of $-8.40 \pm 0.97$ and $-8.74 \pm 1.07$ and these translate to a mean distance of $14.7 \pm 3.8$ kpc, which is adopted as the distance to Nova Sco 2015. The extinction value of $A_V=2.23$ used in these calculations is slightly larger than the total galactic extinction of 1.99 in the direction of the nova as estimated by Schlafly \& Finkneiner (2011) from the dust extinction maps.

\subsection{General properties of the spectra}

\begin{figure}
\centering
\includegraphics[bb=-1 0 316 603,width=3.5in,height=7.0in,clip]{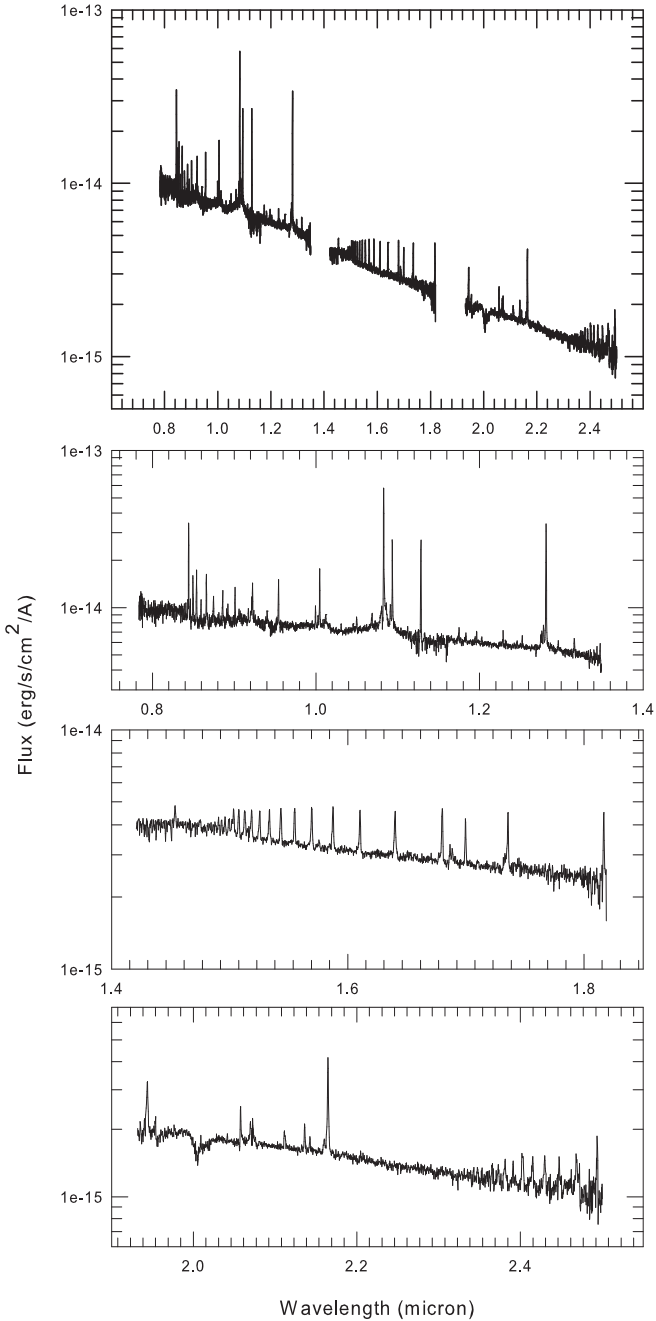}
\caption[]{The entire IRTF spectrum of Nova Sco 2015 obtained on 2015 March 23.625UT covering the 0.8 to 2.5 $\mu$m region is shown in the top panel. Magnified view of the J, H and K sections of the spectrum are shown in the next three panels.}
\label{fig_NSco_IRTF}
\end{figure}

\begin{figure*}
\centering
\includegraphics[bb=0 0 546 219,width=6.25in,height=2.5in,clip]{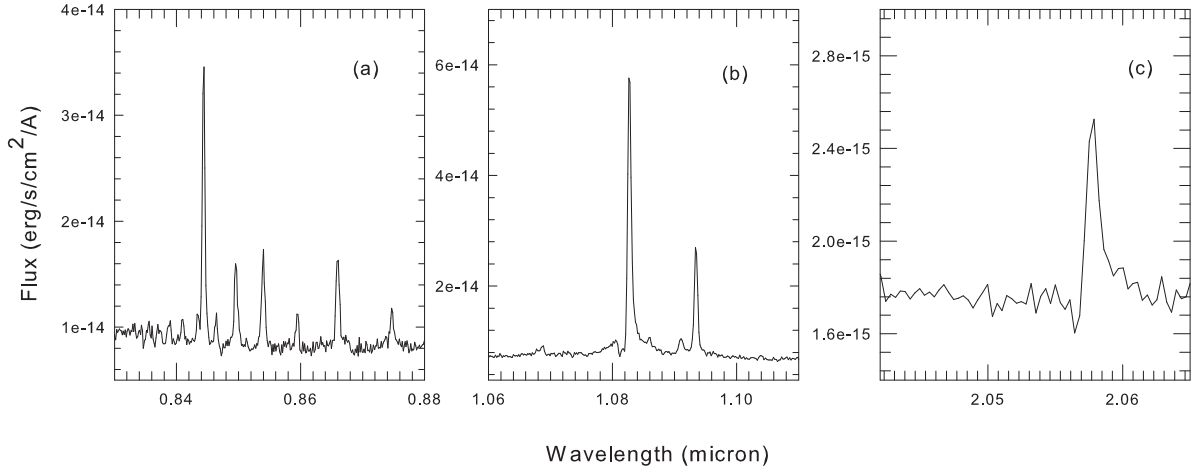}
\caption[]{Magnified sections of some of the lines from the IRTF spectra of Nova Sco 2015. Panel (a) shows the OI 8446 and the NIR  Ca triplet at 8498.023, 8542.091 and  8662.141A respectively. Panel (b) shows the HeI 1.0831 micron line and panel (c) shows the P Cygni feature of the HeI 2.058 micron line}
\label{fig_NSco_IRTF_Magnified}
\end{figure*}

The near-infrared spectra of Nova Sco 2015 at different epochs are shown in Fig~\ref{NSco15_spec_JHK} and Fig~\ref{fig_NSco_IRTF}. The prominent spectral features in these spectra are the Brackett and Paschen recombination lines of H I and He I lines at 1.0831, 1.7002 and 2.0581 $\mu$m, with the 1.0831 $\mu$m line being overwhelmingly strong. The N I blend at 1.2461 and 1.2469 $\mu$m and the Lyman $\beta$ fluoresced O I lines at 0.8446 \& 1.1287 $\mu$m are also present. These spectra are typical of of the He/N class of nova with strong lines of He I seen starting from the first set of observations on 7.16 d after the outburst. The absence of C I lines all through the  span of present observations is also consistent with the He/N class (Banerjee \& Ashok, 2012). The P Cygni absorption features are clearly seen in the higher resolution IRTF spectra obtained on 2015 March 23.625 UT. Another notable feature seen in the IRTF spectra is the presence of blue emission components in the profiles of Pa $\gamma$, Pa $\beta$ and Br $\gamma$ HI emission lines. The magnified sections of the selected lines from the IRTF spectra are shown in Figs~\ref{fig_NSco_IRTF} and ~\ref{fig_NSco_IRTF_Magnified} to highlight the P Cygni features and the weak blue components. A detailed list of emission lines observed in the spectra is given in the Appendix and Table~\ref{table_LineList}.


\par
\subsection{Evidence for a shock from the evolution of the line profiles}
\par
With the help of the present near-IR observations, we establish that the secondary component of Nova Sco 2015 is a late type cool giant star (see section~\ref{subsec_SecondaryNature}). The evolution of the velocity profiles seen in the emission lines of Nova Sco 2015 spectra suggests and supports this possibility. The T CrB sub class of recurrent novae (RNe) with a giant cool red companion typically show a significant decrease in the width of the emission line profiles with time after outburst (Banerjee et al. 2014).  This behavior is  expected as the high velocity ejecta thrown out  during the eruption  moves through the wind of the companion and thereby undergoes a deceleration. Such a deceleration causes a fast temporal decrease of the expansion velocity resulting in the narrowing of the emission line widths. This behavior has been well documented in the NIR in  the case of 4 other similar symbiotic  systems viz. in the  2006 outburst of RS Oph (Das, Banerjee \& Ashok 2006), in V407 Cyg (Munari et al. 2011) where the donor is a high mass losing Mira variable, in the recurrent nova V745 Sco (Banerjee et al. 2014) and in Nova Sco 2014 (Joshi et al. 2015).
\par
\begin{figure}
\centering
\includegraphics[bb= 0 52 288 349,width=3.0in,height=3.0in,clip]{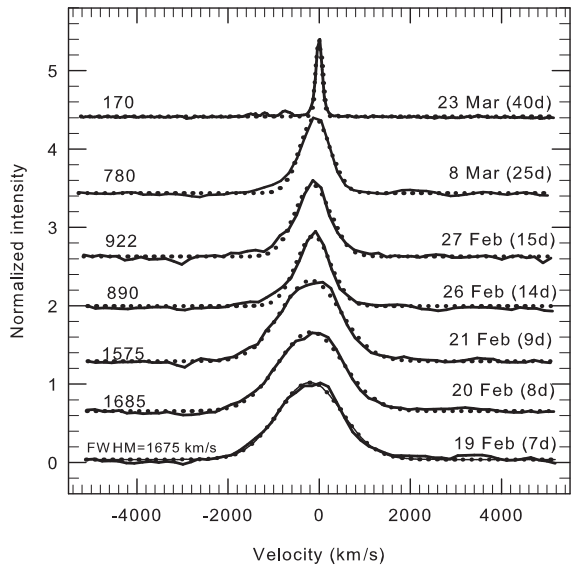}
\caption[]{ Temporal evolution of the Pa$\beta$ 1.2818 $\mu$m line profile in Nova Sco 2015 showing the fast decline in the expansion velocity as the shocked emitting gas decelerates. The measured FWHM (kms$^{-1}$) for each epoch is given along with the velocity profile. The date of observation is indicated as also the days after outburst (in brackets) }
\label{PaBeta_TimeEvol}
\end{figure}
\begin{figure}
\centering
\includegraphics[angle=-90,width=0.50\textwidth]{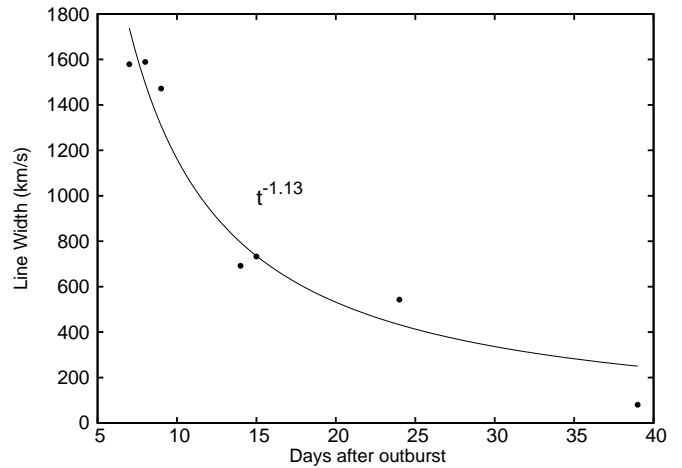}
\caption[]{Time evolution of the line width (FWHM) of Pa$\beta$ 1.2818 $\mu$m line for Nova Sco 2015. The FWHMs have been deconvolved for instrumental broadening; see text for further details.}
\label{fig_NSco_FWHMvsTime}
\end{figure}

Our near IR observations of Nova Sco 2015 show a similar behavior. Fig~\ref{PaBeta_TimeEvol} shows the evolution of the Pa$\beta$ 1.2818 $\mu$m line profile during our observations. The narrowing of the line profile is clearly observed here. Fig~\ref{fig_NSco_FWHMvsTime} shows the time evolution of the observed line widths (FWHM) of the Pa$\beta$ 1.2818 $\mu$m line. The intrinsic FWHM of the profiles have been obtained by deconvolving the  observed profiles from instrumental broadening  by assuming  a gaussian profile for both the observed and instrumental profiles (a reasonable assumption) from which it follows that the  FWHMs will combine in quadrature ($FWHM_{intrinsic}^2$ + $FWHM_{instrument}^2$ = $FWHM_{observed}^2$). The FWHM of the instrumental profile  for 2015 February 19 to 2015 March 8 data from NICS on Mt. Abu Telescope is measured to be 560 kms$^{-1}$. For the 2015 March 23 data from the IRTF Telescope, the same is measured to be  150 kms$^{-1}$ from an argon lamp arc spectrum which is equivalent to the resolution of 2000 cited for SpeX.  A power law fit to the evolving intrinsic line widths, of the form $t^{-\alpha}$, is shown in Fig~\ref{fig_NSco_FWHMvsTime} which  is seen to give a  reasonable fit for a  value of $\alpha = 1.13 \pm  0.17$.

\par
The impact of the high-velocity nova ejecta  with the  wind of the  giant companion is known to produce a strong shock which can heat the gas to high temperatures. This hot, shocked gas  can be the site of  hard X-rays (Sokolski et al. 2006; Bode et al. 2006)  and also $\gamma$-ray production created by diffusive acceleration of particles across the shock to TeV energies. The accelerated protons can subsequently either inverse-Comptonize ambient low-energy radiation to the $\gamma$ ray regime or participate in production of neutral pions  which decay with the emission of  gamma rays. The early X-ray and radio observations of Nova Sco 2015 by Nelson et al (2015)  implicate the presence of such a strong shock forming in the red giant wind. Bode \& Kahn (1985), for e.g., have  discussed the propagation of such a shock wave into the dense ambient medium surrounding the white dwarf. It may be described as a three stage process viz.:
\begin{enumerate}
\item A free expansion or ejecta dominated stage where the ejecta expands freely into the red giant wind. This phase lasts till the mass of the swept-up material from the donor wind is smaller than the mass of the the nova ejecta. A constant velocity of the shock is seen during this time.
\item An adiabatic phase or  Sedov-Taylor stage where the majority of the ejecta kinetic energy has been transferred to the swept-up ambient gas and there is negligible cooling by radiation losses. This phase is characterized by the temporal evolution of shock velocity $v$ as $ \propto t^{\rm 1/3}$, assuming a $r^{\rm -2}$ dependence for the decrease in density of the wind.
\item In phase 3, the shocked material has cooled by radiation, and here the expected dependence of the shock velocity is $ v \propto t^{\rm −1/2}$.
\end{enumerate}

\par
In the case of Nova Sco 2015 the free expansion stage, if it occurred in the first instance,  is clearly missed. This is  possibly because our observations began late viz. our earliest spectrum being recorded 7 days after the outburst. The deceleration that accompanies phases 2 or  3 is seen but the  decay is too fast and the index  $\alpha = 1.13 \pm 0.17 $ that we get  deviates substantially from that expected in either phase 2 or 3. Such deviations were also noticed in other recurrent novae as well e.g. RS Oph (Das et al. 2006), V745 Sco (Banerjee et al. 2014). This likely happens due to the propagation of ejecta into a non-symmetrical wind. In such cases, the ejecta would be slowed down more effectively in the parts moving in the direction of the giant due to the increasing density in that direction. In addition there could be anisotropic distribution of the material over the equatorial plane. Thus as a combination, the mass distribution of the ejecta around the white dwarf would be anisotropic and the shock front would then propagate as an aspherical one (Chomiuk et al. 2012, Fig 6 therein).
\par

\begin{figure}
\centering
\includegraphics[bb= 0 -1 247 297,width=3.0in,height=3.5in,clip]{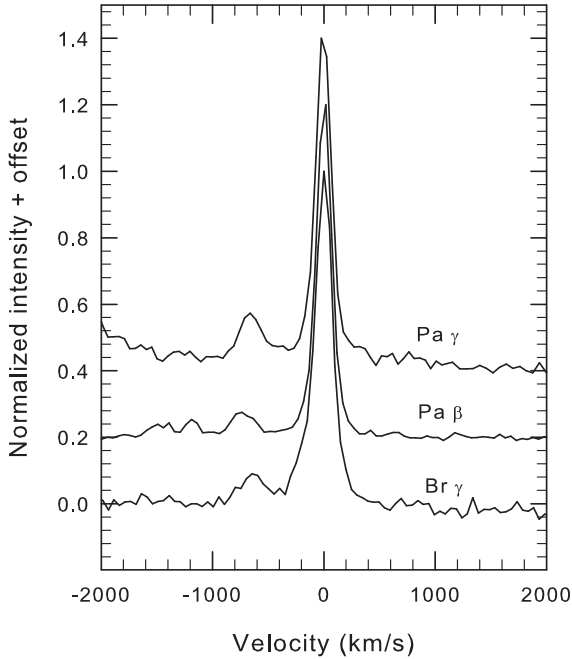}
\caption[]{The velocity profiles of Pa$\gamma$ 1.0938, Pa$\beta$ 1.2818 and the Br $\gamma$ 2.1656 $\mu$m lines. A weak blue component is seen in each of the profiles, separated from the principal profile, by -650, -765, -690 km/s for  Pa$\gamma$, Pa$\beta$  and the Br $\gamma$ lines.}
\label{fig_NSco_VelPlot}
\end{figure}

This nova possibly has a bipolar flow associated with it based on the description of the  early optical spectrum by Walter (2015) where, apart from the main central feature,  symmetrically displaced emission features at about $\pm$ 4500 kms$^{-1}$ were seen in the H$\alpha$ profile. Such a profile structure is typical of a bipolar flow   and has been seen in quite a few novae viz.  RS Oph (Banerjee et al. 2009), KT Eri 2009 (Ribeiro et al , 2013; Raj et al. 2013), T Pyx (Joshi et al. (2014). From our own data,  indication for an asymmetrical ejecta flow also comes from the velocity profiles of the Pa$\gamma$ $1.094\mu$m, Pa$\beta$ $1.2818\mu$m and Br$\gamma$ $2.1656\mu$m lines shown in Fig~\ref{fig_NSco_VelPlot}. Here a weak blue component is seen in each of the profiles, separated from the principal profile, by -650, -765, -690 $kms^{-1}$ for  Pa$\gamma$, Pa$\beta$  and the Br $\gamma$ lines respectively. This could be the blue symmetrically displaced component found by Walter (2015) which has undergone considerable deceleration from its original value of  -4000 km/s down to the present value of $\sim$ -600 to -700 km/s. The intensity of its counterpart red component could have dropped below detection limits. In short,  indications are clearly present for deviations from spherical symmetry in the velocity kinematics seen in Nova Sco 2015. This could be one of the reasons for the deviation of the deceleration index $\alpha$ from model values.


\subsection{Case B recombination line analysis}
\label{subsec_NSco2015_CaseBAnalysis}

\begin{figure}
\centering
\includegraphics[angle=0,width=0.50\textwidth]{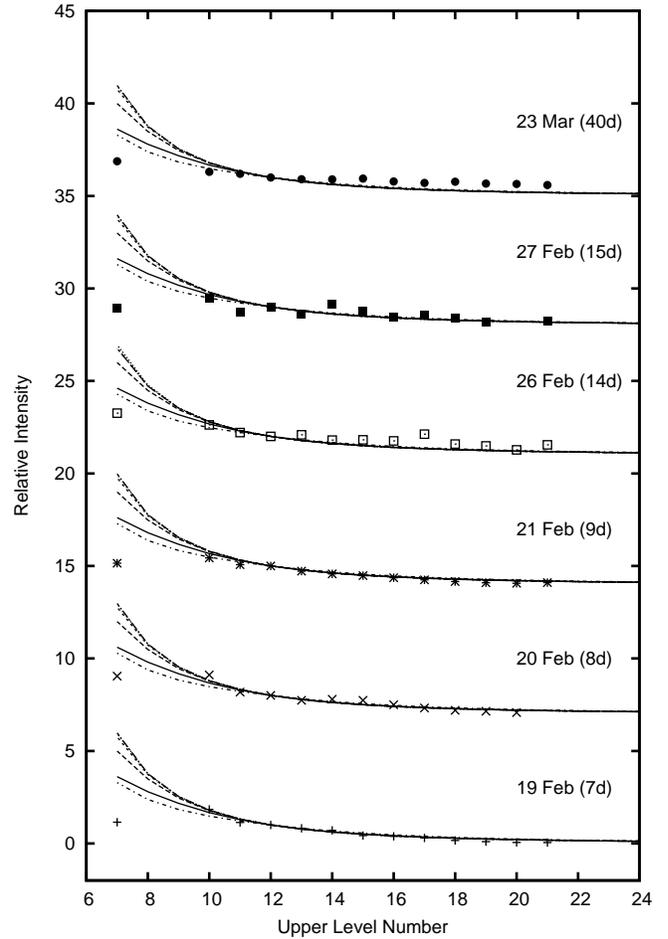}
\caption[]{Recombination Case B analysis for the Hydrogen Brackett series of lines for Nova Sco 2015. The X axis is upper level principle quantum number of the line transition. The line intensities (data points) are normalized with respect to Br 12 line strength and offsets are added for the sake of clarity. The model predictions for Case B analysis are also shown for temperature = 10000 K and electron number densities of
10${^{\rm 9}}$ cm${^{\rm -3}}$ (Small dash-dotted lines),
10${^{\rm 10}}$ cm${^{\rm -3}}$ (solid lines),
10${^{\rm 11}}$ cm${^{\rm -3}}$ (big dash lines),
10${^{\rm 12}}$ cm${^{\rm -3}}$ (small dash lines),
10${^{\rm 13}}$ cm${^{\rm -3}}$ (dotted lines), and
10${^{\rm 14}}$ cm${^{\rm -3}}$ (dash-dotted lines).
}
\label{NSco15_RecombAnalysis}
\end{figure}

We have performed the recombination case B analysis following the same lines as done earlier for Nova Cep 2014. We analyze observed spectra spanning 6 epochs covering the first 33 d of our observations. The measured line strengths from the flux calibrated spectra of N Sco 2015 are given in Table~\ref{table_LineLuminosity}.  Flux calibrations of the spectra are done using near IR photometric observations from SMARTS consortium \footnote {www.astro.sunysb.edu/fwalter/SMARTS/NovaAtlas/nsco2015/nsco2015.html}. Fig~\ref{NSco15_RecombAnalysis} shows the observed relative line strengths for the Brackett lines which have been normalized with respect to  Br12 set to unity.  Br22 and Br23 line values (as given in Table~\ref{table_LineLuminosity}) are not considered for this analysis as they are not resolved properly. The predicted case B values  are shown in Fig~\ref{NSco15_RecombAnalysis} for a representative temperature 10000 K and for electron number densities of $n_e$ = 10$^{9}$, 10$^{10}$, 10$^{11}$, 10$^{12}$,10$^{13}$ and 10$^{14}$ $cm^{\rm -3}$. As can be seen, in this nova also, the Br$\gamma$ line is weaker than expected implying it is optically thick. Thus, using the same formalism described for Nova Cep 2014, the emission measure $n_e^2L$ is estimated to be in the range of $1.3 \times 10^{33}$ to $7.7 \times 10^{33}$ cm$^{-5}$, the corresponding lower limit of the  electron density $n_e$ is in the range $1.8\times 10^9$ cm$^{-3}$ to  $5.8 \times 10^{9}$ cm$^{-3}$ and the  mass $M$ is  between  ($4.5 \times 10^{-6}$ --  $2.6 \times 10^{-4} $) $\epsilon$ M$_{\odot}$  where $\epsilon$ is the filling factor. The electron density and mass estimates are again reasonably consistent with expected values.


\begin{table*}
\centering
\caption{Emission line flux values at different epochs for Nova Sco 2015}
\begin{tabular}{lllclll}
\hline\\

Emission Line  &  & &Line Flux on   & & &\\
and Wavelength & & & days after outburst  & & &\\
($\mu$m)       & & & ($10^{-20}$ W/cm$^2$)  & & &\\
\hline\\
	
	                & 7.16d    & 8.16d    & 9.17d   & 14.18d   & 15.18d   & 40.0d \\	
\hline\\
\hline\\
Pa9 0.9226              & 414.0    & 396.0    & 301.0    & 102.0    & 171.0    &  2.07  \\
Pa8 0.9546              & 158.0    & 182.0    & 145.0    & 41.1     & 22.4     &  1.36  \\
Pa7 1.0049$^a$          & 349.0    & 427.0    & 391.0    & 84.8     & 71.2     &  1.44  \\
HeI + Pa6 $^b$     & 2890.0   & 2970.0   & 2380.0   & 624.0    & 568.0    &  9.79$^c$ \\
OI 1.1287		& 246.0    & 194.0    & 185.0    & 49.5     & 23.8     &  3.31 \\
Pa5 1.2818		& 680.0    & 862.0    & 635.0    & 161.0    & 158.0    &  3.95 \\
Br22+ Br23  $^d$     &  ----    &  ---     & 9.87     &  ----    & 1.65     &  0.37 \\
Br21 1.5133		& 5.62     & ----     & 7.71     & 5.42     & 2.04     &  0.20 \\				
Br20 1.5192		& 5.83     & 6.11     & 4.83     & 2.73     & ----     &  0.22 \\				
Br19 1.5261		& 10.9     & 11.5     & 7.19     & 4.85     & 1.88     &  0.22 \\				
Br18 1.5342		& 17.9     & 15.6     & 12.1     & 5.85     & 3.69     &  0.26 \\			
Br17 1.5439		& 32.9     & 25.9     & 19.3     & 11.2     & 4.81     &  0.23 \\			
Br16 1.5556		& 41.8     & 39.3     & 28.0     & 7.59     & 4.07     &  0.26 \\				
Br15 1.5701		& 46.6     & 58.5     & 36.7     & 8.15     & 6.76     &  0.31 \\				
Br14 1.5881		& 77.0     & 62.9     & 43.6     & 8.00     & 10.2     &  0.30 \\				
Br13 1.6109		& 89.3     & 58.2     & 55.3     & 10.8     & 5.59     &  0.30 \\				
Br12 1.6407		& 109.0    & 79.1     & 76.3     & 9.94     & 9.01     &  0.33 \\				
Br11 1.6807		& 123.0    & 93.2     & 81.9     & 12.0     & 6.56     &  0.40 \\
HeI  1.7002		& 54.5     & 31.3     & 31.0     & 4.70     & 3.92     &  0.19 \\
Br10 1.7362		& 201.0    & 167.0    & 110.0    & 16.1     & 13.4     &  0.44 \\
HeI 2.0581		& 79.7     & 71.2     & 56.8     & 7.10     & ----     &  0.19 \\
HeI 2.112		& 20.3     & 17.5     & 13.2     & 1.62     & ----     &  0.08 \\
Br7 2.1655		& 125.0    & 162.0    & 87.5     & 22.5     & 8.59     &  0.63 \\

\hline
\hline
\end{tabular}
\label{table_LineLuminosity}
\begin{list}{}{}
 \item a : Blended with other lines
 \item b : HeI 1.0831 is blended with HI Pa6 1.0938
 \item c : The integrated line flux of HeI and Pa6
 \item d : Integrated line flux of Br22 and Br 23
\end{list}
\end{table*}



\subsection{The symbiotic nature of the secondary and possibility of recurrent nova}
\label{subsec_SecondaryNature}
\par
Our near-IR observations along with the archival data from 2MASS survey indicate that Nova Sco 2015 presents a strong case to belong to the class of symbiotic system consisting of a WD and and a late type giant companion. As symbiotic systems are rare among novae, identification of a new object likely to belong to this group is of much significance.
\par
The bright near IR counterpart of Nova Sco 2015 from the 2MASS archival database (2MASS J17032617-3504178) is likely to be a symbiotic system based on its  2MASS magnitudes of $J=13.40 \pm 0.03 $, $H=12.53 \pm 0.04$ \& $K=12.22 \pm 0.03$. These magnitudes are transformed to Bessel \& Brett (1988) homogenized system using transformation equations given by Carpenter (2001) and corrected for interstellar extinction using the relations given by Rieke \& Lebofsky (1985). The IR color indices are thus obtained as $(J-H) =  0.69 \pm 0.02$, and $(H-K) = 0.14 \pm 0.02$. These colors are consistent with the values of 0.68 and 0.14 expected for K3 III / K4 III respectively (Bessel \& Brett 1988).
\par
\begin{figure}
\centering
\includegraphics[angle=-90,width=0.50\textwidth]{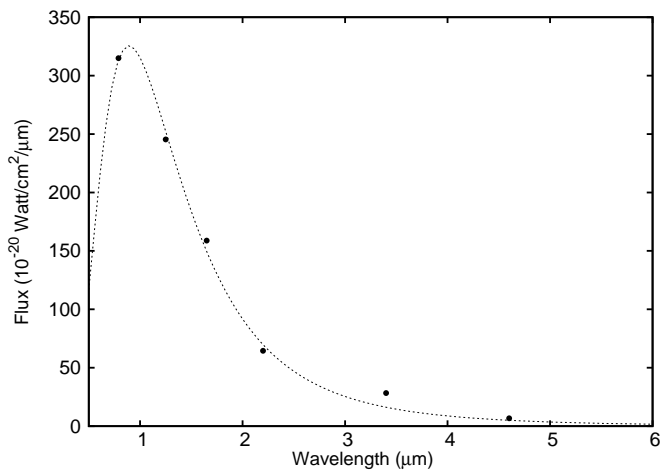}
\caption[]{Spectral Energy Distribution (SED) of the pre-outburst near-IR counterpart of Nova Sco 2015 based on  data  from 2MASS, WISE and DENIS. A blackbody spectrum with a temperature of 3225 $\pm$ 50 K gives a good fit to the spectral energy distribution.}
\label{NSco2015_SED}
\end{figure}

On the other hand the SED in quiescence suggests a  slightly different class for the secondary. For constructing the SED,  the wavelength coverage was extended on the either size of the 2MASS coverage by using WISE W1 and W2 bands data and DENIS I band data. It may be noted that the DENIS magnitudes of $J$ = 13.45 $\pm$ 0.07  and $K$ = 12.20 $\pm$ 0.10   are in good agreement with the corresponding 2MASS values of 13.40 $\pm$ 0.03 and  12.22 $\pm$ 0.03 respectively. The WISE W3 and W4 magnitudes were not used for the SED because the star is not seen as a point source in the WISE images in these bands, only the background diffuse IR cirrus is being picked up. The extremely low SNR (between 3 to 6) and the poor quality flags of the W3 and W4 data show that they should not be used. All magnitudes were corrected for  interstellar extinction using A$_V$ = 2.23. The SED shown in Fig~\ref{NSco2015_SED}, using I(0.79 $\mu$m) = 15.25 $\pm$ 0.04, W1(3.3 $\mu$m) = 11.24 $\pm$ 0.03 and W2(4.6$\mu$m) = 11.46 $\pm$ 0.03, is well fit by a black body of temperature $3225 \pm 50$ K, suggesting a spectral class of M4-5 III. Considering the spectral class of K3 III / K4 III suggested earlier from the near-infrared $JHK$ colors it appears that the determination of the spectral type of the companion is uncertain to few sub-classes $-$ at best we can say that range of the likely spectral class is either K or M. To have a more definitive classification, the spectral lines of the secondary must be recorded in a good SNR spectrum in quiescence.
\par
Further support for the secondary to be in the giant class comes from its absolute magnitude estimation. Assuming that the quiescent K band brightness is dominated by the secondary (i.e m${_K}$ of the secondary  = 12.22 from 2MASS) and using the MMRD distance estimate to the nova of $\sim$ 14.7 kpc, the K band absolute magnitude of the secondary, is calculated as $-3.87$. Whereas using the intricsic color $(V-K)_{intrinsic} = 3.60$ (Bessell \& Brett 1988) and absolute magnitude $M_{V} = -0.20$ (Lang 1990) for K5 III spectral class, the corresponding $M_{K}$ is determined as $-3.80$. As the two are in good agreement, it further strengthens the classification of the secondary as a red giant. On the other hand, the possibility of the companion being a dwarf instead of a giant can be ruled out from the following consideration. For  a mid-K to mid-M spectral class dwarf as suggested by our previous analyses, the K absolute magnitudes is  in the range of $9.2 - 4.2$ respectively (Pecaut \& Mamajek 2013; see online version of table 5 therein). Using the distance modulus relation, the corresponding distance range comes out to be $50 - 400$ pc. This is severely in disagreement with the distance estimated earlier using MMRD relations. Further for such a close distance range of $50 - 400$ pc, the extinction versus distance models of Marshall et al. (2006) suggest that we should get an extinction value of 0.45 for A$_V$ which is again inconsistent with the observed estimate of 2.23.


An additional confirmation for the symbiotic nature of Nova Sco 2015 comes from the comparison of its H$\alpha$ and  R band images. The SUPERCOSMOS archive \footnote{http://www-wfau.roe.ac.uk/sss/H$\alpha$/} H$\alpha$ image is considerably brighter than the R band counter part indicating the presence of strong H$\alpha$ emission.  The SUPERCOSMOS values of H$\alpha$ and short-R band magnitudes are 15.48 and 16.33 respectively, whereas the mean value of the (H$\alpha$ - short R) magnitude  for about 275 listed sources  in a 3 arc-minute square  field  around the object is found to be $-0.29$. That is, the source is considerably bright in H$\alpha$.  In fact, pronounced H I emission is used as one of the  principal criteria for the classification of symbiotic stars (Belczynski et al.,  2000). This can be seen from the spectra of symbiotic stars, for e.g.  in the catalogue of Munari \& Zwitter (2002),  wherein they are seen to exhibit strong  H$\alpha$ emission. However in addition to the presence of HI lines, the other criteria  for a definitive symbiotic star classification requires the presence of  higher excitation lines (e.g. [OIII], HeII). After the system has returned to quiescence, and the nova ejecta faded, it may be checked whether such lines are seen.
\par
We estimate the outburst amplitude of Nova Sco 2015 by associating the optical counter part NOMAD-1 0549-0492872 with $V$ = 17.0 as the progenitor suggested by Guido \& Howes (2015). The $V$ = 9.492 near the maximum from the SMARTS database gives an outburst amplitude (A) of $\sim$7.5 in the $V$ band which is relatively small. It lies $\sim$ 4.5 magnitude below the outburst amplitude (A) vs log ($t_2$) plot for classical novae (Warner 1995; Fig 5.4).
\par
In an extensive study of a large sample of classical novae and the recurrent novae Pagnotta \& Schaefer (2014) have identified the characteristics common to recurrent novae to identify the potential recurrent novae among the known classical novae. The data discussed earlier shows that most of these characteristics, namely, small outburst amplitude, near IR colors resembling the colors of late type giant, expansion velocity exceeding 2000 kms$^{-1}$ and the presence of high excitation lines are fulfilled by Nova Sco 2015, thus it presents a strong case for to be a potential recurrent nova. It is worth noting  the similar case of Nova Sco 2014, wherein Joshi et al (2015) have shown that the outburst occurred in a symbiotic binary system and also suggested that it could be a recurrent nova.

\section{Discussion}
No $\gamma$-ray emission was detected by the FERMI observatory from Nova Sco 2015 or  Nova Sco 2014 (Joshi et al. 2015) which are both similar systems.  On the other hand, two other similar symbiotic systems were detected in $\gamma$ rays viz. V407 Cyg and V745 Sco. The non-detections in Nova Sco 2014 and Nova Sco 2015 could be a  consequence of both objects being sufficiently  distant that any emission from them falls below the  detection threshold of FERMI. But it is desirable to check the FERMI data  from both these novae carefully for any weak or suggested signs of detection. This has bearings on the origin of $\gamma$-ray emission from novae where the latest paradigm suggests that $\gamma$-ray emission could be a generic property of all novae and not intrinsic to just symbiotic systems (Ackermann et al. 2014). Novae from which $\gamma$-ray emission has been detected, but which are not symbiotic systems, are  Nova Sco 2012, Nova Mon 2012,  Nova Del 2013 and Nova Cen 2013.


\section{Summary}
We present near-infrared photometric and spectroscopic observations of Nova Cep 2014 and Nova  Sco 2015 which were discovered in outburst on  2014 March 8.79 UT and 2015 February 11.84 UT respectively. Our observations for Nova Cep 2014 cover 9 epochs from 5 to 90 days after outburst and for Nova Sco 2015 cover 11 epochs covering 7 to 47 days after outburst. Nova Cep 2014 shows the conventional characteristics of a Fe II class characterized by strong CI lines together with HI and O I lines, whereas Nova Sco 2015 is classified as He/N class, shows strong He I emission lines together with HI and OI emission features. Using MMRD relations for the novae, we estimate the distances for Nova Cep 2014 and Nova Sco 2015 as $17.2 \pm 7$ kpc and $14.7 \pm 3.8 $ kpc respectively. For Nova Sco 2015, the presence of a decelerative shock seen through a narrowing of the line profiles,  presents a strong case for it to be a symbiotic system. We discuss the evolution of the strength and shape of the emission line profiles. The ejecta velocity shows a power law decay with time ($t^{-1.13 \pm 0.17}$) and case is presented for asymmetric ejecta flow in the winds of a cool giant companion star. The SED of the secondary in quiescence shows it to be a late cool type giant and the H$\alpha$ excess seen from the system in quiescence is also indicative of the symbiotic nature of the system. Constraints are put on the spectral type of the companion star. A Case B recombination analysis shows the  Brackett lines to be optically thick. This in turn helps us  to estimate, for Nova Sco 2015, that the emission measure $n_e^2L$ is  in the range of $1.3 \times 10^{33}$ to $7.7 \times 10^{33}$ cm$^{-5}$, the corresponding lower limit of the  electron density $n_e$ is in the range $1.8\times 10^9$ cm$^{-3}$ to  $5.8 \times 10^{9}$ cm$^{-3}$ and the  mass $M$ of the ejecta is  between  ($4.5 \times  10^{-6}$ --  $2.6 \times 10^{-4}$)$\epsilon$ M$_{\odot}$ where $\epsilon$ is the filling factor. For Nova Cep 2014, the corresponding estimates are $1.3 \times 10^{33}$ to $7.7 \times 10^{33}$ cm$^{-5}$ for the emission measure, $2.2\times 10^9$ cm$^{-3}$ to  $1.1 \times 10^{10}$ cm$^{-3}$ for the electron density and  ($5.4 \times 10^{-6}$ --  $1.3 \times 10^{-4}$)$\epsilon$  M$_{\odot}$ for the mass of the ejecta where $\epsilon$ is the filling factor.


\section{Acknowledgments}
 The research work at the Physical Research Laboratory is funded by the Department of Space, Government of India. We acknowledge the use of data from the AAVSO, SMARTS, 2MASS, WISE and DENIS databases. We wish to thank V. Venkataraman and Vishal Joshi for taking some of the observations of Nova Cep 2014 from Mt. Abu. Helpful comments from the anonymous referee are acknowledged with thanks. DS is a visiting astronomer at the Infrared Telescope Facility which is operated by the University of Hawaii under contract NNH14CK55B with the National Aeronautics and Space Administration. DS would like to thank Michael Cushing for his efforts in keeping the Spectool reduction code up to date.



\appendix
\section{Line Identification}

Most of the NIR lines that appear in the spectra of novae have been identified in Das et al (2008). However, the spectra presented there were in the 1.08 to 2.4 $\mu$m  region, whereas the present spectra are taken with a newer instrument extend up to 0.85 $\mu$m. A robust identification of the numerous lines that appear in the 0.85 to 1.08 $\mu$m (IJ band region) is thus desirable. To identify the lines that contribute to a nova's spectrum, we use an  LTE model to build  synthetic spectra as in Das et al (2008) and Ashok \& Banerjee (2003).
\par
Assumptions of LTE may not strictly prevail in an nova environment although, around maximum and the early decline stage, when the particle density can be high (up to even $10^{14} cm^{-3}$) collisions will be a dominant mechanism and will tend to drive the gas towards a Boltzmann distribution and LTE. Yet, in spite of the limitations of the LTE assumption we find that the model-generated spectra, greatly aid in a  more secure identification of the lines observed. Briefly (more details in Das et al, 2008) the model spectra are generated by considering only those elements whose lines can be expected at discernible strength. Since nucleosynthesis calculations of elemental abundances in novae (Starrfield et al. 1997; Jose $\&$ Hernanz, 1998) show that H, He, C, O, N, Ne, Mg, Na, Al, Si, P, S  are the elements with significant yields in novae ejecta, only these elements have been considered. The Saha ionization equation was applied to calculate the fractional percentage of the species in different ionization stages and subsequently the Boltzman equation was applied to calculate level populations. By switching off or greatly increasing the abundance of an element, it is easy to identify the positions where the lines of that element disappear or build up.
\par
Fig~\ref{fig_NCep_LineID} shows the IJ band spectrum of Nova Cep 2014 of 2014 March 20 in black and a typical synthetic LTE spectrum in gray below. The LTE spectrum has been computed for $n_e$ =  10${^{\rm 9}}$ cm${^{\rm -3}}$, $T$ = 8000K and abundances  typically found in CO novae as given in Starrfield et al. (1997) and Jose $\&$ Hernanz (1998). A total of $\sim$ 2500 of the strongest lines  were considered for these elements compiled from the Kurucz atomic line list{\footnote {http://cfa-www.harvard.edu/amp/ampdata/}} and National Institute of Standards and Technology (NIST){\footnote {http://physics.nist.gov/PhysRefData/ASD}} line list database. Based on the line identifications done here, and lines in novae spectra known from earlier studies (Williams, 2012; Das et al. 2008), the observed line list is given in the Table~\ref{table_LineList}.

\begin{table*}
\label{table_LineList}
\begin{center}
\caption{A list of emission lines identified from the $JHK$ spectra of Nova Cep 2014 and Nova Sco 2015}
\begin{tabular}{lrrrrlrrrr}
\hline\\
Wavelength & Species & Other con- & Nova Cep & Nova Sco &Wavelength & Species & Other con- & Nova Cep & Nova Sco        \\
(${\mu}$m) &   & tributors/ & & & (${\mu}$m)& & tributors/& &               \\
 &   & Remarks& & & & & Remarks& &               \\
\hline
\hline
0.8359	 &	HI Pa22	 &		 &		 &	x	 &	1.2527	 &	He I	 &		 &		 &	x	 \\
0.8374	 &	HI Pa 21	 &		 &		 &	x	 &	1.2562,1.2569	 &	C I 	 &		 &	x	 &		 \\
0.8392	 &	HI Pa20	 &		 &		 &	x	 &	1.2620	 &	C I 	 &		 &	x	 &		 \\
0.8413	 &	HI Pa 19	 &		 &		 &	x	 &	1.2755	 &	u.i	 &		 &		 &	x	 \\
0.8446	 &	O I	 &	 Pa18 0.8438	 &		 &	x	 &	1.2763	 &	u.i	 &		 &		 &	x	 \\
0.8467	 &	Pa17	 &		 &		 &	x	 &	1.2818	 &	HI Pa5	 &		 &	x	 &	x	 \\
0.8498	 &	Ca II	 &	 Pa16  0.8502	 &		 &	x	 &	1.2963	 &	u.i	 &		 &		 &	x	 \\
0.8542	 &	Ca II	 &	 Pa15 0.8545	 &		 &	x	 &	1.3164	 &	O I	 &		 &	x	 &	x	 \\
0.8598	 &	HI Pa14	 &		 &		 &	x	 &	1.34-1.38	 &	N I  	 &	Blend of many 	 &	x	 &		 \\
 &		 &		 &		 &		 &		 &	 	 &	 NI lines	 &	 	 &		 \\
0.8665	 &	HI Pa13	 &	Ca II 0.8662	 &		 &	x	 &	1.4420	 &	C I	 &		 &	x	 &		 \\
0.8680	 &	NI 	 &		 &	x	 &		 &	1.4543	 &	C I	 &		 &	x	 &		\\
0.8750	 &	HI Pa12	 &		 &		 &	x	 &	1.4539	 &	u.i	 &	CI 1.4543?	 &		 &	x	 \\
0.8802	 &	u.i	 &	0.8807 Mg I?	 &		 &	x	 &	1.4757	 &	N I	 &		 &	x	 &		 \\
0.8863	 &	HI Pa11	 &		 &		 &	x	 &	1.4906	 &	HI Br27	 &		 &		 &	x	 \\
0.8909	 &	u.i	 &		 &		 &	x	 &	1.4938	 &	HI Br26	 &		 &		 &	x	 \\
0.8923	 &	u.i	 &		 &		 &	x	 &	1.4967	 &	HI Br25	 &		 &		 &	x	 \\
0.9015	 &	HI Pa10	 &		 &		 &	x	 &	1.5000	 &	HI Br24	 &		 &		 &	x	 \\
0.9021	 &	NI 	 &		 &	x	 &		 &	1.5039	 &	HI Br23	 &		 &		 &	x	\\
0.9089	 &	CI 	 &		 &	x	 &		 &	1.5083	 &	HI Br22	 &		 &		 &	x	\\
0.9174	 &	u.i	 &		 &		 &	x	 &	1.5133	 &	H1 Br21	 &		 &		 &	x	 \\
0.9226	 &	HI Pa9	 &		 &		 &	x	 &	1.5192	 &	HI Br20	 &		 &	x	 &	x	 \\
0.9264	 &	OI 	 &		 &	x	 &		 &	1.5261	 &	HI Br19	 &		 &	x	 &	x	\\
0.9396	 &	NI 	 &		 &	x	 &		 &	1.5342	 &	HI Br18	 &		 &	x	 &	x	\\
0.9406	 &	CI 	 &		 &	x	 &		 &	1.5439	 &	HI Br17	 &		 &	x	 &	x	\\
0.9402	 &	u.i	 &	 CI 0.9406?	 &		 &	x	 &	1.5556	 &	HI Br16	 &		 &	x	 &	x	 \\
0.9546	 &	HI Pa8	 &		 &		 &	x	 &	1.5701	 &	HI Br15	 &		 &	x	 &	x	 \\
0.9863, 0.9872	 &	NI 	 &		 &	x	 &		 &	1.5881	 &	HI Br14	 &		 &	x	 &	x	 \\
0.9993	 &	u.i	 &		 &		 &	x	 &	1.6005	 &	C I	 &		 &	x	 &		 \\
1.0049	 &	HI Pa7	 &		 &		 &	x	 &	1.6109	 &	HI Br13	 &		 &	x	 &	x	 \\
1.0112	 &	NI 	 &	CI 1.0119	 &	x	 &		 &	1.6407	 &	HI Br12	 &		 &	x	 &	x	 \\
1.0124	 &	He II 	 &		 &		 &	x	 &	1.6807	 &	HI Br11	 &		 &	x	 &	x	 \\
1.0308	 &	u.i	 &		 &		 &	x	 &	1.6872	 &	Fe II 	 &		 &		 &	x	 \\
1.0399	 &	u.i 	 &		 &	x	 &		 &	1.6890	 &	C I	 &		 &	x	 &	x	 \\
1.0457	 &	u.i	 &		 &		 &	x	 &	1.7002	 &	He I	 &		 &		 &	x	 \\
1.0497	 &	u.i	 &		 &		 &	x	 &	1.7362	 &	HI Br10	 &		 &	x	 &	x	 \\
1.0534	 &	NI 	 &		 &	x	 &		 &	1.7200-1.7900	 &	C I 	 &	Blend of many	 &	x	 &		 \\
	 &		 &		 &		 &		 &		 &		 &	 CI lines	 &	  &		\\
1.0685	 &	C I	 &		 &	x	 &	x	 &	1.7413	 &	FeII	 &		 &		 &	x	 \\
1.0831	 &	He I	 &		 &	x	 &	x	 &	1.8174	 &	HI Br9	 &		 &		 &	x	 \\
1.0938	 &	HI Pa6	 &		 &	x	 &	x	 &	1.9446	 &	HI Br8	 &		 &		 &	x	 \\
1.1287	 &	O I	 &		 &	x	 &	x	 &	1.9722	 &	C I 	 &		 &	x	 &		 \\
1.1330	&	C I	&		&	x	&		&	2.0581	 &	He I	 &		 &	x	 &	x	 \\
1.1659	&	C I	&		&	x	&		&	2.0703	 &	u.i	 &		 &		 &	x	 \\
1.1753	 &	C I	 &		 &	x	 &	x	 &	2.1023	 &	 C I	 &		 &	x	 &		\\
1.1828	 &	Mg I	 &		 &		 &	x	 &	2.1120, 21132	 &	He I	 &		 &		 &	x	\\
1.1880,1.1896	 &	C I	 &		 &	x	 &	x	 &	2.1156-2.1295	 &	CI 	 &	Blend of 	 &	x	 &		 \\
	 &		 &		 &		 &		 &		 &		 &	CI lines	 &		 &		 \\
1.1969	 &	He I	 &		 &		 &	x	 &	2.1361	 &	u.i	 &		 &		 &	x	 \\
1.2028	 &	u.i	 &		 &		 &	x	 &	2.1425	 &	u.i	 &		 &		 &	x	\\
1.2249,1.2264	 &	C I 	 &		 &	x	 &		 &	2.1655	 &	HI Br7	 &		 &	x	 &	x	\\
1.2281	 &	u.i	 &		 &		 &	x	 &	2.2906	 &	C I	 &		 &	x	 &		\\
1.2461, 1.2469	 &	N I	 &		 &	x	 &	x	 &	2.3438 - 2.4945	 &	HI Pf30 to 17	 &		 &		 &	x	 \\
\hline
\hline
\end{tabular}
\end{center}
\end{table*}


\end{document}